**Low-Voltage Printed, All-Polymer Integrated Circuits Employing a Low-leakage and High-Yield Polymer Dielectric**

*Elena Stucchi, Giorgio Dell'Erba, Paolo Colpani, Yun-Hi Kim and Mario Caironi\**

E. Stucchi, Dr. G. Dell'Erba, P. Colpani, Dr. M. Caironi
Center for Nano Science and Technology @PoliMi, Istituto Italiano di Tecnologia, via Pascoli 70/3, 20133 Milano, Italy
E-mail: mario.caironi@iit.it

E. Stucchi
Dipartimento di Fisica, Politecnico di Milano, P.za Leonardo da Vinci 32, 20133 Milano, Italy

Prof. Y.-H. Kim
Department of Chemistry and Research Institute of Green Energy Convergence Technology (RIGET), Gyeongsang National University, Jinju 660-701, Republic of Korea

In the path toward the integration of organic field effect transistors (OFETs) and logic circuits into low-cost and mass produced consumer products, all-organic devices based on printed semiconductors are one of the best options to meet stringent costs requirements. Within this framework, it is still challenging to achieve low voltage operation, as required by the use of thin film batteries and energy harvesters, for which a high capacitance and reliable organic dielectric is required. Here, the development of a parylene-C based dielectric bilayer compatible with top-gate architectures for low-voltage OFETs and logic circuits is presented. The polymer bilayer dielectric allowed the high yield fabrication of all-polymer, bendable, transparent *p*- and *n*-type OFETs operating below 2 V, with low leakage, uniform performances and high yield. Such result is a key enabler for the reliable realization of complementary logic circuits that can operate already at a voltage bias of 2 V, such as well-balanced inverters, ring-oscillators and D-Flip-Flops.

## 1. Introduction

Organic field effect transistors (OFETs) have been extensively studied in the last decades, and are currently considered as the building block of the next generation of thin film electronics.[1-3] This technology allows for the development of large area, inexpensive, transparent electronic devices by means of cost-effective manufacturing techniques. Organic materials can be processed from solution, mostly by means of printing and coating techniques, which are low-temperature processes, enable mass production and minimize the by-products.[4] Being these manufacturing techniques compatible with flexible, plastic and other low-cost substrates, there is a clear potential for the integration of additional electronic functionalities into mass-produced, consumer products.[5-7] In order to meet very stringent costs constraints, which do not allow integration of conventional electronic chips, all-organic circuits fabricated by means of scalable techniques are one of the best options.

One of the main limitations hindering the adoption of all-organic printed circuits in real application is related to their operating voltage.[8] In order to grant their portability and easy integration, these circuits need in fact to be powered by thin film batteries[9, 10] and/or energy harvesters, such as plastic solar cells,[11, 12] thus requiring maximum operation voltages of a few volts and low power consumption, while keeping reasonably high values of accumulated charge density and of current flowing into the circuits. Efficient low voltage operation can be achieved by acting on the capacitance of the dielectric layer, which should be as high as possible and at the same time guarantee optimal charge accumulation and transport at the semiconductor-dielectric interface of both holes and electrons, in order to enable complementary architectures to drastically reduce power consumption.[13, 14] Many efforts have been recently devoted to the development of suitable polymer materials for gate dielectric applications, aiming at the achievement of the highest possible gate capacitance.[13] Two

WILEY-VCHmain strategies may be followed, either increasing the dielectric constant of the employed material or decreasing its thickness. The integration of high-*k* materials as dielectrics is not straightforward, as the energetic disorder at the interface might interfere with charge transport inside the semiconductor layer.[15, 16] Multilayer structures combining low-*k* and high-*k* materials have been therefore introduced.[17, 18] However, high-*k* materials typically show dielectric relaxations occurring at low frequency,[19, 20] possibly limiting the maximum operation frequency of OFETs. Such limit is particularly severe in electrolyte gated transistors,[21] where huge capacitances are achieved at the expense of the switching speed because of ions motion, even in recent solid-state electrolytes.[22] To avoid such limitations, an obvious choice is to downscale a low-*k* dielectric polymer. However, this strategy typically leads to very high leakage currents or complete breakdown of devices. A possible solution to the latter is the adoption of poly(chloro-p-xylene)-C (Parylene-C), a low-*k*, semicrystalline, thermoplastic polymer, which is commonly deposited by means of chemical vapor deposition (CVD);[23] this is not a solution-based printing technique, but it has already been shown to be large-area compatible and industrially scalable, so it would not be a limiting factor during a possible upscale of the manufacturing. This dielectric grants a low energetic disorder at the semiconductor-dielectric interface[24] and presents some appealing characteristics, such as its chemical inertness, flexibility, and the ability to form conformal, pinhole-free coatings.[24, 25] For such reasons, parylene has been largely investigated as a dielectric for OFETs, although most of the reports deal with bottom-gated structures,[26-33] where the dielectric is deposited on an inert surface and not on the active material. Staggered, top-gate structures are however very important towards the development of logic circuits as they allow for optimal charge injection,[34] thanks to the possibility of reducing the contact resistance through gate to source/drain electrodes overlap.[35] Moreover, they allow for controlled interfaces between the semiconductor and the dielectric layer,[36, 37] and for a partial self-encapsulation effect thanks to the upper gate dielectric and the gate electrode which improves environmental stability of devices.[38] In the context of OFETs based on a printed semiconductor, only one example of top-gated device employing parylene as dielectric has been reported, with opaque evaporated metallic contacts and operating voltages in the order of 30 V, thus not compatible with low voltage operation.[39]

Here we exploit a thin parylene layer for the development of a high capacitance polymer bilayer dielectric, comprising an ultra-thin poly(methyl methacrylate) (PMMA), compatible with top-gate OFETs. Our bilayer dielectric achieves areal capacitances in the order of 20 nF cm$^{-2}$ with leakage currents below 1 nA cm$^{-2}$. It enables the robust fabrication of low-voltage, transparent, fully-polymeric, and complementary OFETs, where all the layers are inkjet printed but the parylene layer, based on poly([N,N'-bis(2-octyldodecyl)-naphthalene-1,4,5,8-bis(dicarboximide)-2,6-diyl]-alt-5,5'-(2,2'-bitthiophene)) (P(NDI2OD-T2) and poly[2,5-bis(7-decylnonadecyl)pyrrolo[3,4-c]pyrrole-1,4(2H,5H)-dione-(E)-1,2-di(2,2′-bithiophen-5-yl)ethene] (29-DPP-TVT) as *n*- and *p*-type semiconductors, respectively. The ideal dielectric characteristics allow the transistors to operate with sub-10 V voltages and low leakage and to be fabricated with a 99 % yield and good uniformity, as probed on a 100 transistors array for each OFET polarity. The reliable process was key to enable the integration of OFETs into logic circuits of different level of complexity operating down to 2 V, from well-balanced complementary inverters to 7-stages ring oscillators with stage delays as low as 1.14 ms, and D-Flip-Flops, fundamental building blocks of registers, counters and timers. Our results, exploiting scalable deposition processes, therefore offer a viable path for the cost-effective integration of electronic functionalities into consumer products.

## 2. Results and discussion

For the realization of high capacitance all-polymer dielectrics, we opted for a bilayer dielectric combining a thin parylene film on top of an ultra-thin solution-processed layer of poly(methyl methacrylate) (PMMA), thus combining the excellent dielectric properties of parylene[40] and the optimal interface formed by PMMA with a broad range of polymer semiconductors. The PMMA layer is in fact important to achieve optimal *n*-type devices (**Figure S1**), as it protects the electron transporting semiconductor from the degrading effect of the chlorine atoms present in the parylene.[39] The overall dielectric properties of the bilayer dielectric have been investigated in order to find the best combination to be integrated into printed OFETs and circuits. For this analysis, a MIM (metal-insulator-metal) structure has been employed, as shown in **Figure 1**a, with the dielectric bilayer deposited between two metallic electrodes. The PMMA layer is 20 nm thick, while for the parylene layer we considered different thicknesses; for the analysis presented here, three different capacitors have been examined, with parylene layers that are 120, 200 and 310 nm thick. The capacitance as a function of frequency for the three above mentioned capacitors has been measured and is shown in Figure 1b. The areal capacitance at 1 kHz is 11.54 nF cm$^{-2}$ for the 310 nm thick parylene layer, and it increases to 16.69 and 19.00 nF cm$^{-2}$ for the 200 nm and 120 nm thick layers, respectively. The average leakage current at 10 V amounts to 10 nA cm$^{-2}$ for the 120 nm thick parylene layer, while the 200 and 310 nm thick layers show average leakage currents of 4.6 and 2.25 nA cm$^{-2}$, respectively (Figure 1c). Bilayers employing parylene layers thinner than ~120 nm, despite leading to an increase in capacitance, typically resulted in defective capacitors or short circuits. Following the capacitors characterization, we therefore decided to adopt bilayers with parylene films with thicknesses of about 100 nm for the printed transistors, so that a good compromise could be reached between low operating voltage, favored by the high capacitance value, and low leakages, together with a high yield and uniform performances.

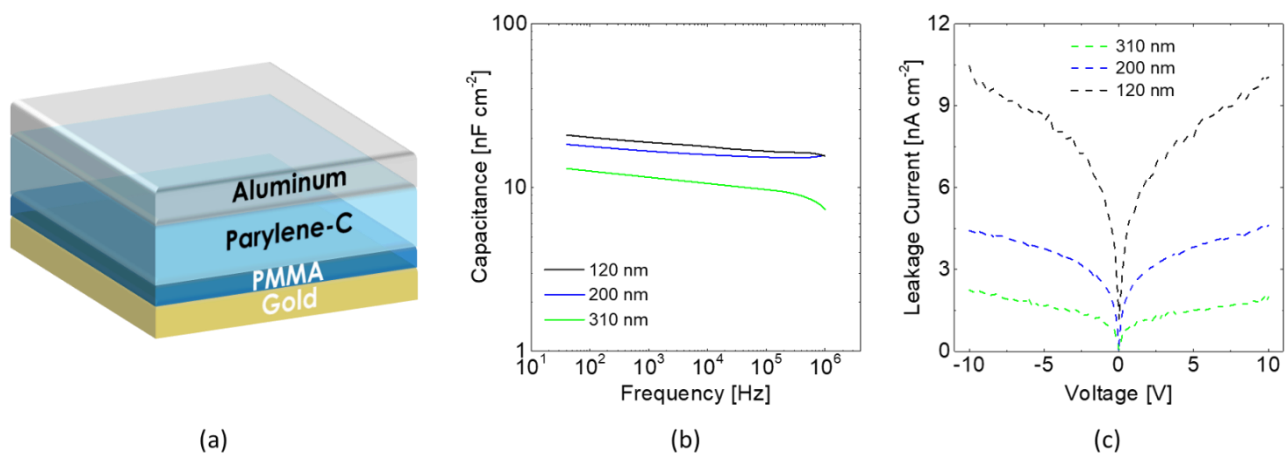

(a)    (b)    (c)

**Figure 1.** (a) Schematic diagram of the MIM structure. (b) Capacitance as a function of the frequency and (c) leakage current density of the capacitors. All the values shown here have been obtained as average over 3 identical devices.

The dielectric bilayer has been integrated into *p*- and *n*-type bottom-contact/top-gated OFETs (**Figure 2**a). These devices have been fabricated exclusively by means of additive processes, which are cost-effective and industrially scalable. First, poly(3,4-ethylenedioxithiophene):polystyrene sulphonate (PEDOT:PSS) source and drain contacts have been inkjet-printed (Figure 2b) onto 125 µm thick polyethylene naphthalate (PEN) substrates, defining a channel length (*L*) of 65 µm and width (*W*) of 1000 µm. Small pads of silver nanoparticles-based ink have been printed in correspondence of the points where electrical contacts are created during the characterization of transistors and circuits; it should be pointed out that these pads are only needed to facilitate external electrical probing of the devices, which can operate also without them, in a fully transparent configuration. Onto the contacts for the *n*-type transistors, a polyethylenimine (PEI)-based injection layer has been printed (Figure 2c), leading to enhanced electrons injection thanks to a reduced electrode work-function and interface doping.[41, 42] P(NDI2OD-T2) has been inkjet printed for *n*-type devices, while 29-DPP-TVT for the *p*-type ones (Figure 2d,e). The dielectric stack has been subsequently deposited on top of the semiconductor: the 20 nm thick PMMA layer has been spin-coated, while parylene has been deposited by means of chemical vapor deposition (CVD), thus obtaining a dielectric stack with an overall thickness lower than 150 nm. Lastly, PEDOT:PSS gate electrodes have been inkjet-printed on top of the dielectric (Figure 2f).

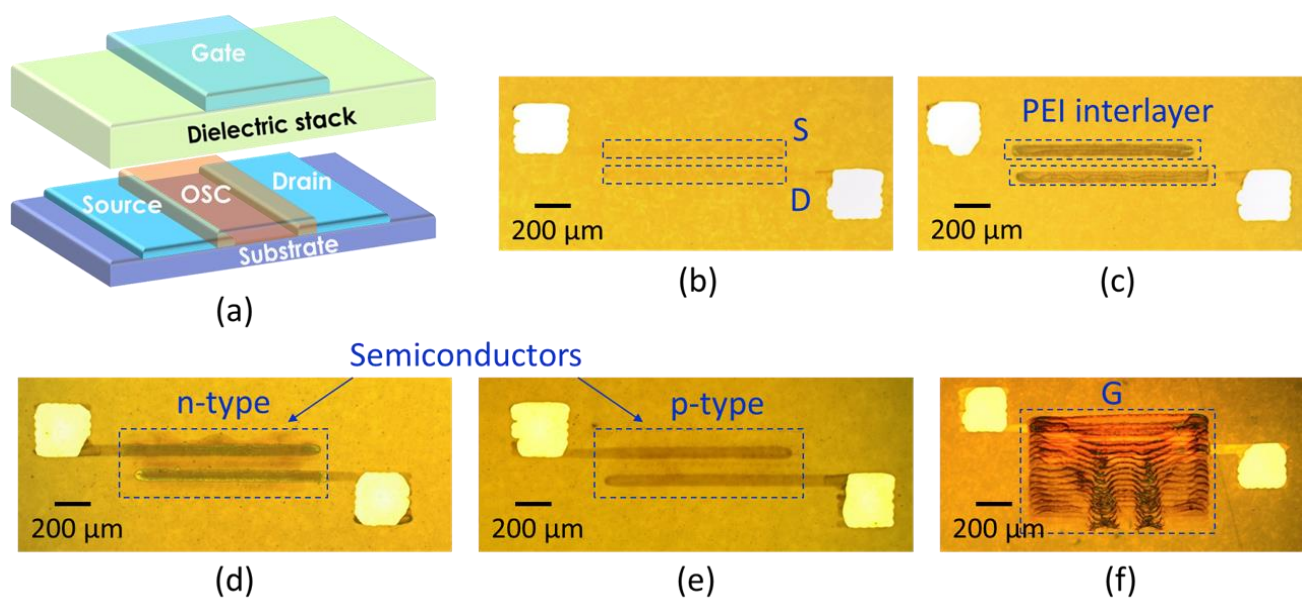

**Figure 2.** Scheme of the printed OFET (a) and optical micrographs of the fabrication steps: (b) PEDOT:PSS source and drain electrodes, together with the silver pads, inkjet printed on the PEN substrate, (c) PEI interlayer inkjet printed on the contacts of the n-type transistors, (d) and (e) inkjet printed P(NDI2OD-T2) and 29-DPP-TVT semiconductors for *n*- and *p*-type devices, respectively, (f) PEDOT:PSS gate electrode, inkjet printed after the deposition of the dielectric bilayer.

The electrical characterization of the resulting OFETs is presented in **Figure 3**. As it can be seen from the transfer curves (Figure 3a,d), both *p*- and *n*-type transistors show proper current modulation at voltages as low as 1 V. Output curves (Figure 3b,e) are rather ideal in the case of *p*-type devices, while a residual S-shaped characteristic is visible for *n*-type devices, likely owing to contact effects, though not precluding a correct low voltage operation of the devices.

The presented electrical data have been extracted from an array of 100 transistors for each polarity, in order to address both the performances and the uniformity of the printed OFETs. The average transfer curves with their standard deviation, for an applied source-drain voltage $V_{DS}$ of 10 V, are shown in Figure 3c,f, while the raw data, both for linear and saturation regime, can be found in **Figure S3**. The average maximum current, which has been extracted from the 10 V transfer curves, amounts to 1.67 ± 0.29 µA for the *n*-type devices, while for the *p*-type devices it amounts to 4.02 ± 0.28 µA. This difference in the performance can be explained by considering the combined effect of their average field effect mobilities and of their threshold voltages.

The average field effect mobility of electrons ($\mu_e$) and holes ($\mu_h$), over 100 devices in both cases, was extracted from the slope of the transfer characteristic curves in the $|V_G|$ range from 8 to 10 V, according to the gradual channel approximation.[43] The areal capacitance for the dielectric stack employed in these devices has been measured for frequencies down to 5 Hz (**Figure S2**c), where we recorded a capacitance equal to 22.6 nF cm$^{-2}$; this value has been used for the mobility extraction. For *n*-type devices, $\mu_e = 0.08$ cm$^2$ V$^{-1}$ s$^{-1}$ in the linear regime ($V_{DS} = 2$ V), and $\mu_e = 0.12$ cm$^2$ V$^{-1}$ s$^{-1}$ in the saturation regime ($V_{DS} = 10$ V), while for *p*-type devices $\mu_h = 0.20$ cm$^2$ V$^{-1}$ s$^{-1}$ in the linear regime, and $\mu_h = 0.21$ cm$^2$ V$^{-1}$ s$^{-1}$ in the saturation one. In the latter case, the mobility values do not show substantial gate voltage dependence in full accumulation (**Figure S4**b). *n*-type OFETs, on the other hand, exhibit a slightly non-ideal behavior, given by a small gate dependence of the mobility curves (Figure S4a). We extracted the measurement reliability factors ($r$), as recommended by Choi et al.,[44] and obtained in both cases high values: close to 100 % for holes mobility in both linear (96 %) and saturation (95 %), and higher than 70 % for electrons mobility (81 % in linear and 73 % in saturation regime). As a result, average effective mobilities ($\mu_{eff}$), which are more robust figure of merits to describe carriers field-effect mobility in thin-film transistors and are calculated as $\mu_h \times r$ and $\mu_e \times r$, are very close to $\mu_h$ and $\mu_e$: for *p*-type devices, $\mu_{eff,lin} = 0.19$ cm$^2$ V$^{-1}$ s$^{-1}$ and $\mu_{eff,sat} = 0.20$ cm$^2$ V$^{-1}$ s$^{-1}$, and for *n*-type devices $\mu_{eff,lin} = 0.06$ cm$^2$ V$^{-1}$ s$^{-1}$ and $\mu_{eff,sat} = 0.08$ cm$^2$ V$^{-1}$ s$^{-1}$. The lower *r* for *n*-type OFETs is due to the combined effect of a weak gate voltage dependence and a higher threshold voltage: P(NDI2OD-T2) devices present a threshold voltage of 1.35 ± 0.41 V and a turn-on voltage of 1 ± 0.22 V, while 29-DPP-TVT transistors show lower current onset values, with a threshold voltage of 0.49 ± 0.21 V and 0.68 ± 0.3 V as turn-on voltage. The subthreshold slope (*SS*) is 148 ± 23 mV dec$^{-1}$ and 126 ± 25 mV dec$^{-1}$ for *p*- and *n*-type OFETs, respectively. On-Off ratios, defined as the maximum ratio of the drain current values in the "on" and "off" states, are relatively high, with average values of 5x10$^5$ for the *p*-type transistors and of 7x10$^6$ for the *n*-type ones.

The transfer curves for devices of both polarities present a small hysteresis. For what concerns *p*-type devices, during the backward sweep, the threshold voltage slightly shifts toward positive voltage and the subthreshold slope (*SS*) is increased. The *n*-type OFETs, on the other hand, present a shift toward negative voltages and the same increase in *SS*. This feature can be explained with the presence of shallow traps at the dielectric interface, which are filled during the forward sweep and remain so during the backward one, which thus appears to be slightly more ideal. Being this hysteresis limited in size and reproducible, it does not affect the performances of our devices in a significant way and can be considered to be negligible.

Overall, of the 200 transistors fabricated for this electrical analysis, only two were not working, leading to a 99% yield. From a simple inspection with the optical microscope of the defective devices, we found that failure in the two devices was caused by particulate that landed onto the active channel (**Figure S5**a,b). It is therefore reasonable to expect that in a more controlled environment, such as a classified cleanroom, the yield may approach 100%.

Gate leakage currents of our transistors are very low, with a flat curve and leakage current values below 100 pA (Figure 3a,d). More than 90% of the transistors are characterized by such low leakage, as shown in the raw data in Figure S3, while only about 5 devices for each polarity present an increased leakage current, while still working properly, with current and mobility values comparable with the ones of the very low leakage devices. For what concerns the transistor arrays presented here, the presence of the leakage current is to be related in most of the cases to the accidental presence of small droplets of the silver nanoparticles ink into the transistors channel, as shown in Figure S5c. Silver pads are needed uniquely for the characterization of transistors and circuits with external probes, but they are not needed for the actual operation of these devices; the main outcome of this work is in fact the development of fully transparent integrated circuits, in which the main cause of defective behavior here highlighted is eliminated.

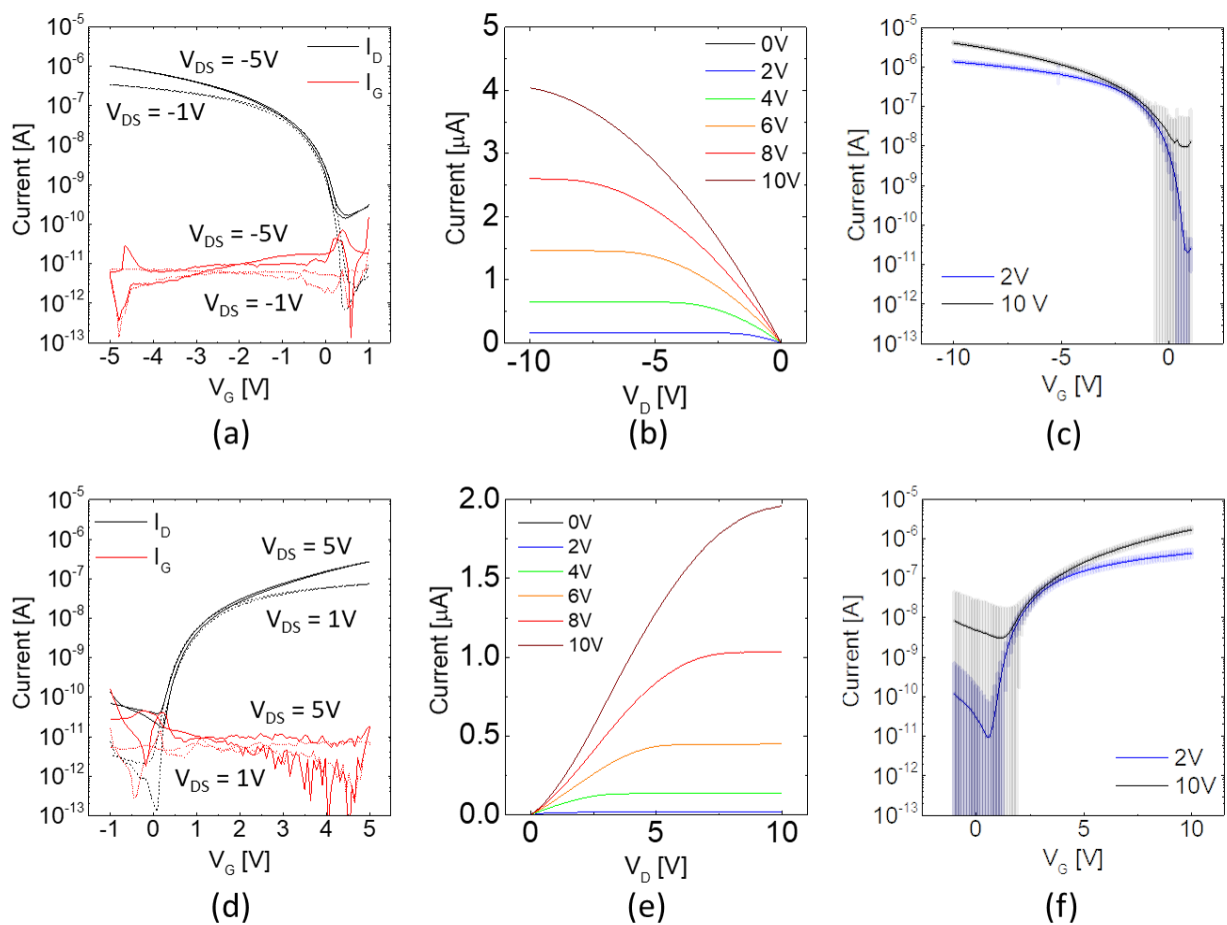

**Figure 3.** Electrical characterization of the printed all-polymeric OFETs ($L = 65$ µm, $W = 1000$ µm). Transfer and output curves for *p*-type (a, b) and *n*-type (d, e) devices, based on 29-DPP-TVT and P(NDI2OD-T2) respectively. Average transfer curves and their standard deviation, for the 10x10 OFETs *p*-type (c) and *n*-type (f) arrays.

The bendability of the printed transistors has been assessed, by means of bending tests with different applied bending radii, equal to 5.5 mm, 8 mm and 13 mm, which lead to strain values equal to about 0.5 %, 0.8 % and 1.1 %. For each bending radius, the transfer curve of one transistor has been measured before the test and after 1, 10, 100 and 1000 bending cycles; the results of this analysis are presented in **Figure 4**. For what concerns the *p*-type devices, there is only a slight increase in the maximum current for increasing number of bending cycles, which is mirrored in the mobility values shown in **Figure S6**a; overall, there is no significant change in performances after 1000 bending cycles. Considering *n*-type transistors, maximum current values stay almost constant up to 1000 bending cycles for applied strains below 1 %, while for a 1.1 % strain value the current and mobility values are unchanged up to 100 bending cycles, while there is a small loss in performances after 1000 cycles, as it can be seen in Figure 4 and Figure S6b.

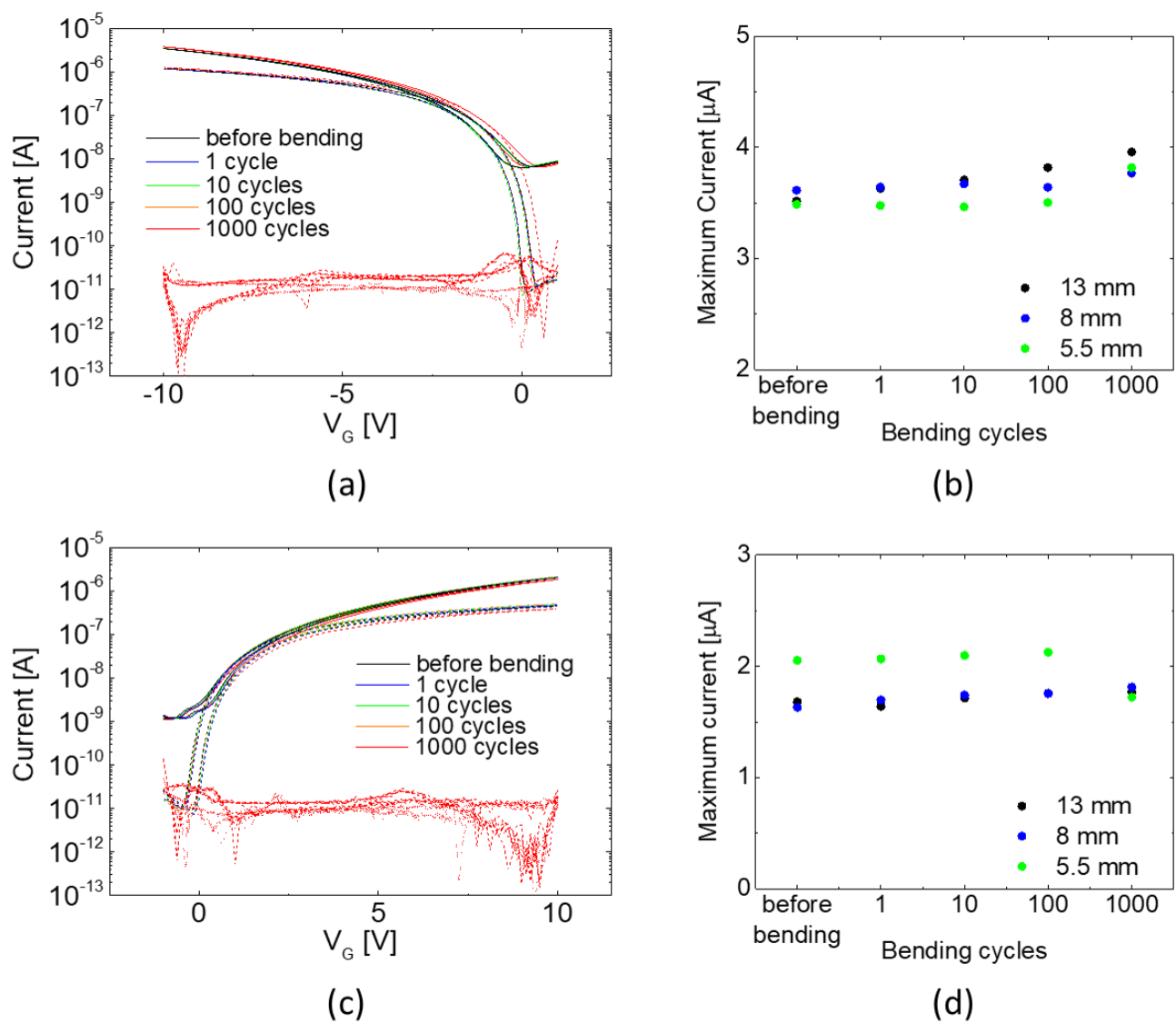

**Figure 4.** Bendability tests results. Transfer curves and maximum current values, recorded before bending and in flat configuration after 1, 10, 100 and 1000 bending cycles, for *p*-type (a, b) and *n*-type (c, d) devices. Transfer curves are presented for the case with an applied strain higher than 1%.

The presented *p*- and *n*-type OFETs have been fabricated with the same techniques on the same substrates, employing identical electrodes and dielectrics. Therefore, their integration into complementary logic circuits is highly simplified. We started by fabricating a logic inverter, the simplest complementary circuit that can be realized, by integrating one *p*- and one *n*-type device. The inverter is fabricated by printing the two transistors with a shared drain contact and the same gate electrode, which act as output and input nodes, following the same fabrication steps presented for the transistor preparation (**Figure 5**a, b). Since the performances of the two different types of devices are not perfectly balanced, with the *p*-type transistors presenting slightly higher currents compared to the *n*-type ones, the channel width of the *p*-type devices ($W_p$) has been designed to be about one third of the *n*-type transistors' one ($W_n$), with $W_p = 300$ μm and $W_n = 1000$ μm, while keeping the same channel length, $L_p = L_n = 65$ μm. The latter dimensions have been adopted for all complementary circuits in this work. First, a static characterization of the printed inverters has been performed. Voltage transfer curves (VTC) have been measured, sweeping the voltage input signal from a "0" logic state to a "1"; supply voltages ($V_D$) varying from 2 V to 10 V have been used, and the so obtained VTC are plotted in Figure 5c as an average over the performances of 5 printed inverters.[45] By finding the intersection point between the VTC and the bisector of the axes, an average inverting threshold voltage of 4.90 V has been obtained for devices operated at 10 V, with a standard deviation of 0.05 V. The inverters are functional down to a supply voltage of only 2 V, for which the average inverting threshold voltage amounts to 1.19 V, with a standard deviation of 0.07 V. The average gain values, obtained as the derivative of the VTC (Figure 5d), amount to about 14 for $V_D = 10$ V, with slightly higher values recorded for smaller input voltages. The maximum recorded gain is 17, and has been achieved with an input voltage of 2 V. The noise margins (*NM*), which give an estimation of the immunity of the printed inverters to noise on the input signal, have been calculated using the maximum equal criteria.[46] *NM* lay in the range between 50 % and 60 % of $V_D/2$ for all supply voltages from 2 to 10 V, with the highest value obtained for $V_D = 2$ V, for which the average *NM* is equal to 0.62 V. The obtained figures of merit are ideal, indicating that low voltage complementary logic gates based on our printed transistors can be used in order to realize more complex integrated circuits.[45] The voltage transfer characteristics of these inverters show a small hysteresis, which can be related to the hysteresis shown by the single transistors and has already been addressed. This phenomenon is limited and reproducible, and thus doesn't affect their correct operation.

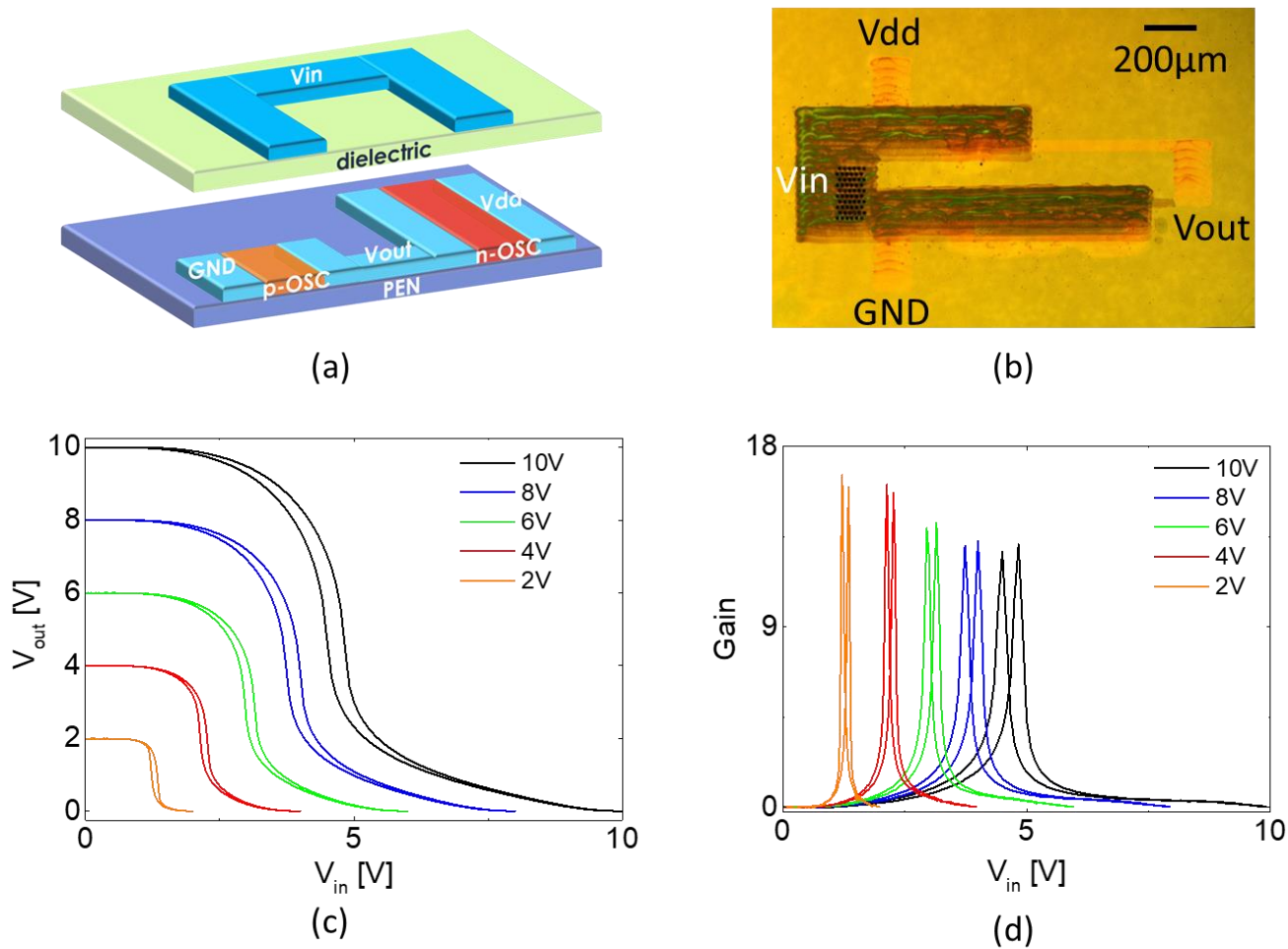

**Figure 5.** (a) Schematic representation and (b) optical micrograph of the printed inverter. (c) Voltage transfer curves and (d) gain curves of the inverters with input voltage varying from 2 V to 10 V.

In order to perform the dynamic characterization of these inverters, 7-stage ring oscillators (RO) have been fabricated (**Figure 6**a). An RO is a loop containing an odd number of inverters, where the output of an inverter is fed into the subsequent one, leading to its characteristic self-oscillation, whose frequency $f_{RO}$ depends on the number of stages ($N$) and on the stage delay ($SD$), according to

$$f_{RO} = \frac{1}{2*N*SD} \quad (1)$$

The oscillating behavior of the circuit has been characterized, with supply voltages varying from 2 V to 10 V, and results for 3 different supply voltages are presented in Figure 6c. The average oscillating frequency, evaluated on 3 different samples, goes from almost 6 Hz at 2 V to 57 Hz for a supply voltage of 10 V, leading to average stage delays of 13.1 ms and about 1.3 ms for supply voltages of 2 and 10 V, respectively (Figure 6b). The best performing circuit oscillates with a frequency of 62.5 Hz when a voltage supply of 10 V is applied, which leads a minimum stage delay achieved of 1.14 ms. At the best of our knowledge, this is the first demonstration of a printed, all-polymeric, transparent ring oscillator operating at low voltage.[8, 47-50]

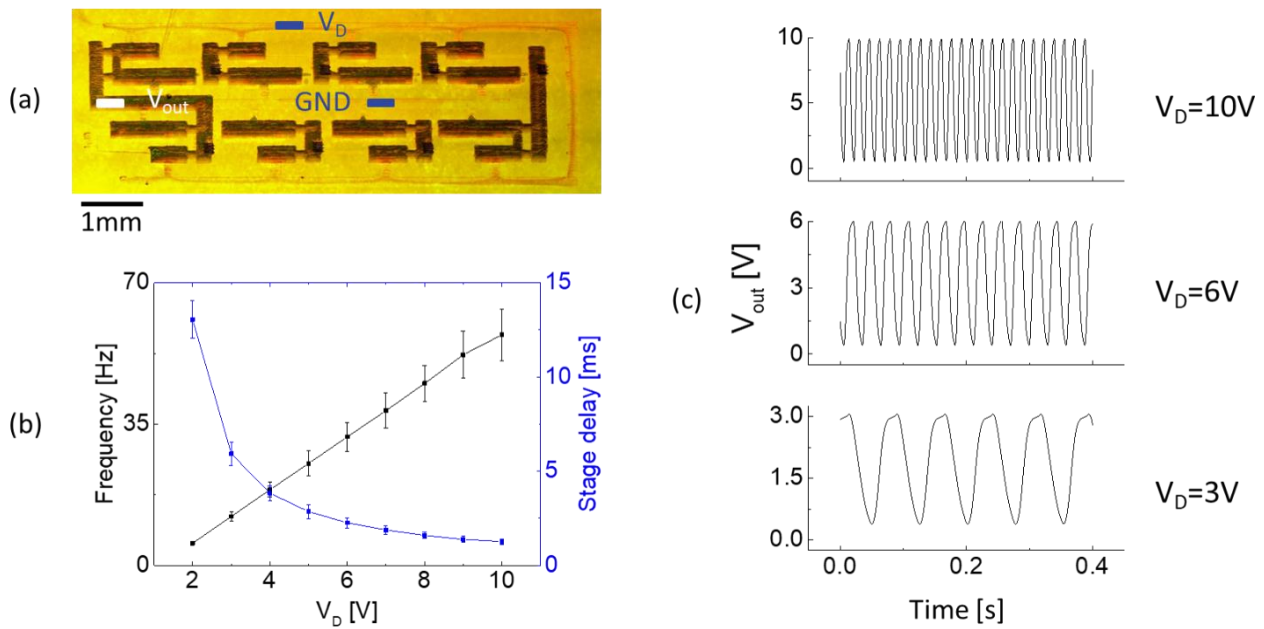

**Figure 6.** (a) Optical micrograph of the printed 7-stage ring oscillator. Shown are the points were electrical connections are created during characterization, namely $V_D$ for the supply voltage, GND for the ground and $V_{out}$ for the output signal. (b) Average oscillating frequency and stage delay, evaluated over 3 different samples, for voltage supply varying from 2 to 10 V. (c) Time response of the 7-stage ring oscillator.

We then proceeded to the fabrication of more complex logic circuitry, in particular we integrated the printed transistor in a D-Flip-Flop (DFF), one of the basic memory elements needed to build sequential logic circuits, counters and timers.[45, 51-54] This DFF has been implemented using a master-slave configuration, in which two D-latches (details in SI) are connected in series with opposite transparency phases. During its operation, this circuit retains the input data, and it adjusts the output according to the stored state only in correspondence of the falling edge of the clock signal; further details about the working mechanism can be found in the supplementary information. For the circuit design we selected a pass-transistor based logic. This logic requires two main building blocks, the inverter, whose development and characterization has already been addressed, and the transmission gate. The latter is a logic port acting as a switch, and in its typical design it is made of two complementary transistors with shared source and drain contacts, and with opposite signals driving the gate electrodes. In our design, aiming at a reduction of the number of transistors used in each circuit, we decided to use a single *p*-type transistor as transmission gate, instead of the complementary port, also because in this case there is no net gain in adopting the complementary pass gate.

As a consequence, our modified DFF design requires only 12 transistors. The schematic and the optical micrograph of our low-voltage transparent DFF are shown in **Figure 7**a,b. In Figure 7d,e the proper operational behavior of our devices is presented, for $V_D$ values equal to 10 V (d) and 2 V (e) (the operation of the D-latches sub-units is shown in **Figure S8**b,c). One of the characteristic feature of these devices is their transparency. We have thus evaluated the transmittance of the printed circuits by means of UV-vis spectroscopy in the visible range, and we have achieved a transmittance equal or higher than 90 % for the overall system; in **Figure S9** we show the spot on the circuit where this measurement has been performed. Taking into account the contribution of the PEN substrate, and normalizing the transmittance for the circuit alone, we can show that our printed circuits have a transparency equal or higher than 95 % in the visible range, as it can be seen in Figure 7c. While previous low-voltage printed DFF has been presented in literature,[55-57] this is the first example, to the best of our knowledge, of a low-voltage, all-polymer, transparent DFF, additionally obtained combining scalable processes.

We assessed the shelf-life stability of non-encapsulated printed circuits stored in nitrogen. After 4 months of storage the DFF are still perfectly operational (Figure 7f), thanks to the fact that characteristics of single transistors and inverters (**Figure S7**) are negligibly altered.

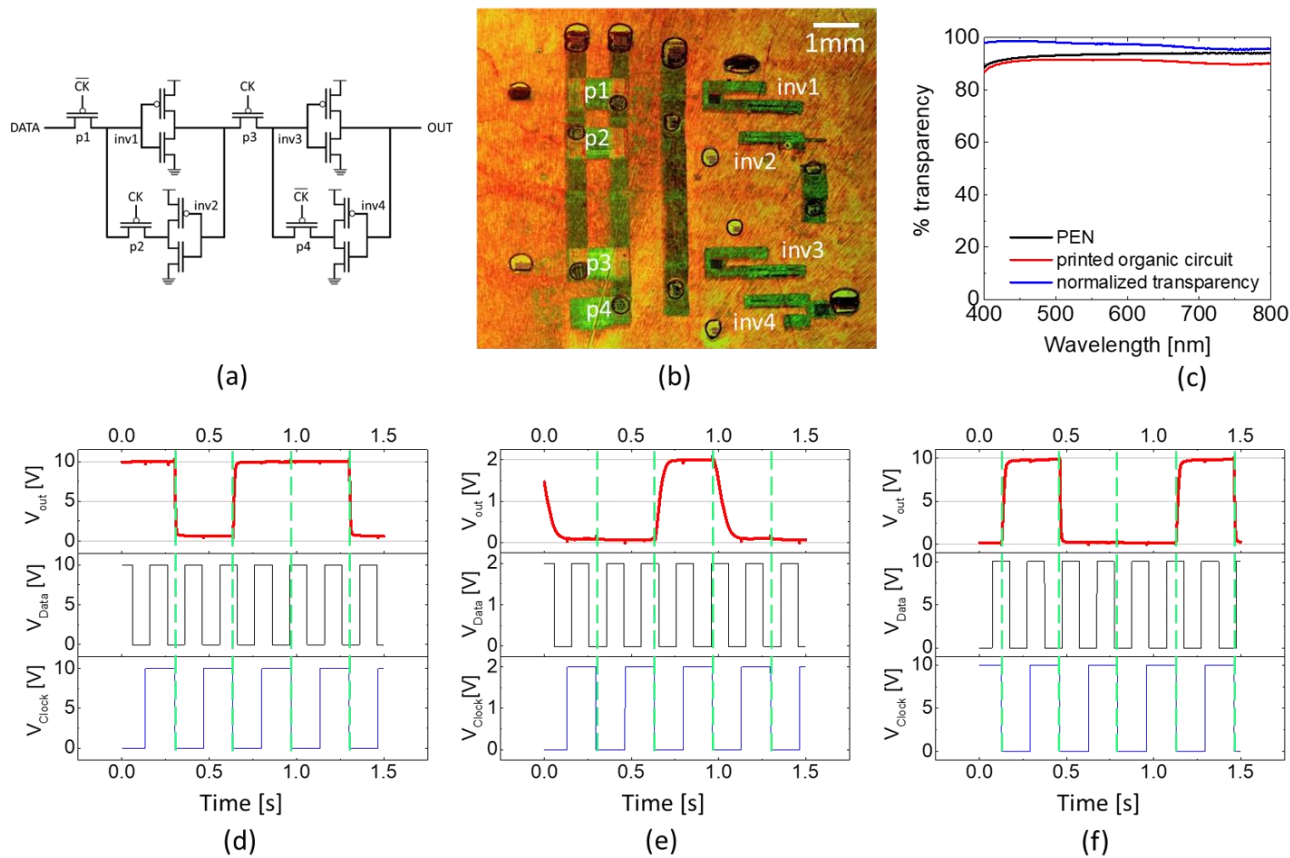

**Figure 7.** Schematic representation (a) and optical micrograph (b) of the transparent, all-polymer D-Flip-Flop. (c) Transparency of the printed DFF with respect to the PEN substrate. Operation of the DFF, for supply voltages of 10 V (d) and 2 V (e). (f) Operation of the same DFF, after 4 months from fabrication.

## 3. Conclusions

We have demonstrated the integration of a parylene-based bilayer as top-gate dielectric for all-polymer, printed, transparent and low-voltage OFETs. Except for the dielectric, in which parylene was deposited through an already industrially scaled chemical vapor deposition method, all the functional layers were inkjet printed on a PEN substrate. We used P(NDI2OD-T2) and 29-DPP-TVT semiconducting co-polymers for *n*- and *p*-type transistors, respectively, and we have reported correct field-effect behavior down to 1 V. The printed OFETs achieve currents higher than 1 µA in both cases for operating voltages in the order of 10 V, with a high yield (99%) and uniform performances. The flexibility of these devices was assessed through bending tests, and it has been shown that they can sustain up to 1000 bending cycles with strains up to 1%.

We successfully demonstrated the integration of these transistors into complementary logic circuits. First, complementary inverters were fabricated, and well balanced output characteristics were obtained, with NM equal or higher than 50 % and an average gain of 14. Next, we demonstrated the first all-polymer, transparent, printed ring oscillators operating at voltages as low as 3 V, and achieving a maximum oscillation frequency of 62.5 Hz for a supply voltage of 10 V, thus showing a stage delay of 1.14 ms. Finally, we demonstrated all-polymer, transparent, printed D-Latches and D-Flip-Flops, correctly operating at voltages as low as 2 V. The transparency of these circuits is higher than 90% for the whole device has been obtained. Life-shelf stability in nitrogen of non-encapsulated circuits exceeds 4 months after fabrication.

Our results show that it is possible to fabricate fundamental logic electronics blocks, as required in serialization circuits, timers and counters, through all-polymer, transparent OFETs on plastic capable of operating at voltages compatible with thin film batteries or energy harvesters, paving the way to their integration into consumer products with low additional costs.

## 4. Experimental Section

Bottom-contact/top-gate organic field effect transistors, whose structure is presented in Figure 2a, have been fabricated on a 125 µm-thick poly(ethylene 2,6-naphthalate) (PEN) substrate, purchased from Du Pont. Poly(3,4-ethylenedioxithiophene):polystyrene sulphonate (Clevios PJ700 formulation, purchased from Heraeus) source and drain contacts (Figure 2b) have been inkjet-printed by means of a Fujifilm Dimatix DMP2831. This process allows to fabricate transistors with a channel width of about 1000 μm and channel length in the order of 65 μm.

In order to facilitate the characterization of these devices, small pads of a silver nanoparticles-based ink have been printed (Silverjet DGP-40LT-15C, purchased from Advanced Nano Products (ANP)) in correspondence of the points where electrical contacts with external probes are made during measurements. This is mainly related to the fact that these devices are transparent and it is thus quite difficult to characterize them without the help of clearly visible contact pads. All the devices presented in this work have been fabricated with silver pads, but it is necessary to point out that they are not required for the actual operation of any of the circuits presented here.

For what concerns the *n*-type transistors, a PEI-based injection layer has been printed on top of the contacts (Figure 2c). The printed solution contains ethylene glycol (20% vol, purchased Sigma Aldrich), a solution of zinc oxide (ZnO) nanoparticles in isopropanol (nanoparticles concentration 2.5% wt, this solution amounts to 30% vol of the total volume of the final solution) and PEI (branched, purchased from Sigma Aldrich, average $M_w \approx 10\,000$), dissolved in water with a weight concentration of 0.2% (50% vol of the final solution).

As semiconductors, P(NDI2OD-T2) (ActivInk N2200, Flexterra Corporation) has been used for the *n*-type OFETs, and 29-DPP-TVT (synthesized according to Yu et al.[58]) for *p*-type devices. Both semiconductors were patterned by inkjet-printing (Figure 2d,e). P(NDI2OD-T2) has been printed from a mesitylene-based solution, with a concentration of 7 mg ml$^{-1}$, while 1,2-dichlorobenzene was used as solvent for 29-DPP-TVT, to yield a concentration of 2.5 mg ml$^{-1}$. After printing, annealing in nitrogen atmosphere have been performed; P(NDI2OD-T2) has been annealed for 3 hours at 120 °C, while 29-DPP-TVT for 5 minutes at 110 °C.

The bilayer dielectric stack is composed of a thin poly(methyl methacrylate) (PMMA) layer (purchased from Sigma Aldrich, average $M_w \approx 120\,000$), and a thicker parylene-C layer (dimer purchased from Specialty Coating Systems). The thickness of the PMMA layer is in the order of 20 nm, while the parylene-C layer is about 100 nm thick. The PMMA layer has been spin-coated and annealed at 100 °C for one hour in a nitrogen atmosphere, while parylene-C is deposited by means of chemical vapor deposition, using a SCS Labcoater 2 – PDS2010 system. Parylene dimer underwent pyrolysis inside the furnace chamber at about 690 °C, while deposition has been performed at room temperature, with a pressure in the chamber in the order of 15 mTorr. PEDOT:PSS gate contacts have been inkjet printed on top of the dielectric layer (Figure 2f). For what concerns circuits, via-holes contacting different conductive layers have been fabricated by either by laser or chemical drilling. The characterization of the transistors and circuits has been performed in an inert nitrogen atmosphere. The transfer and output curves for the transistors, together with the voltage transfer curves for the inverters, have been measured using an Agilent B1500A Semiconductor Parameter Analyzer. The dynamic characterization of the integrated circuits has been performed using the above mentioned Semiconductor Parameter Analyzer for recording the output signals, while the input signals and the voltage supply were provided by an Agilent 81150A waveform generator and an Agilent B2912A source meter. For what concerns the dielectric characterization, an Agilent 4294A impedance analyzer has been employed.

**Supporting Information** Supporting Information is available from the Wiley Online Library or from the author.

**Acknowledgements**

We would like to acknowledge Filippo Melloni for his help with the electrical characterization. This work would not have been possible without the support by the European Research Council (ERC) under the European Union's Horizon 2020 research and innovation program "HEROIC", Grant Agreement No. 638059. This work has also been partially supported by the Italian Ministry of Economic Development through the project F/050241/03/X32 under the "Fondo per la Crescita Sostenibile" – Call HORIZON2020 PON I&C 2014-2020. The authors are thankful to the company FLEX for very useful discussions.

Received: ((will be filled in by the editorial staff))
Revised: ((will be filled in by the editorial staff))
Published online: ((will be filled in by the editorial staff))

**A parylene-based high-capacitance dielectric bilayer enables high-yield fabrication of low-voltage printed polymer integrated circuits on plastic,** based on top-gate field effect transistors with low-leakage and high-uniformity. All-polymer complementary circuits, such as 7-stages ring oscillators and D-Flip-Flops, properly operating at voltages as low as 2 V, are demonstrated.

**Keyword** organic field-effect transistors, printed integrated circuits, parylene, inkjet printing, flexible electronics

E. Stucchi, G. Dell'Erba, P. Colpani, Y.-H. Kim, M. Caironi*

**Low-Voltage Printed, All-Polymer Integrated Circuits Employing a Low-leakage and High-Yield Polymer Dielectric**

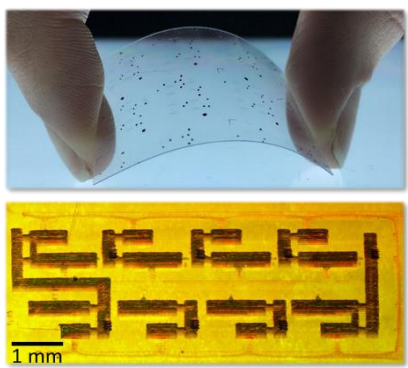

Supporting Information

**Low-Voltage Printed, All-Polymer Integrated Circuits Employing a Low-leakage and High-Yield Polymer Dielectric**

*Elena Stucchi, Giorgio Dell'Erba, Paolo Colpani, Yun-Hi Kim and Mario Caironi*

**Role of PMMA interlayer**

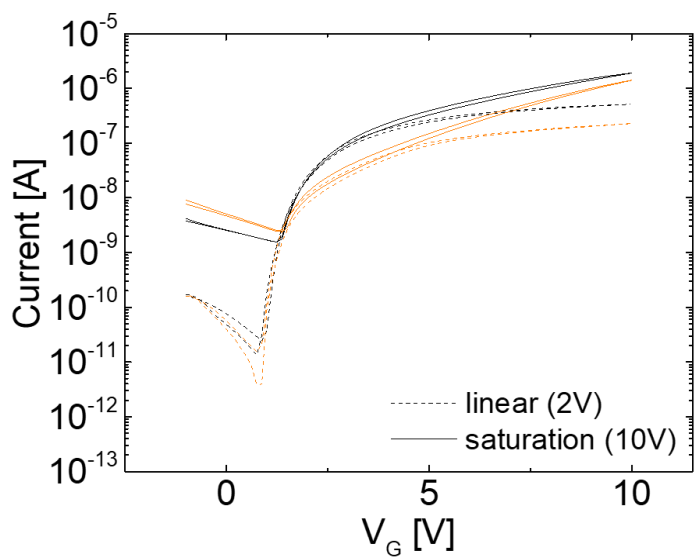

**Figure S1.** Transfer curves of two transistors employing identical structures except for the dielectric layers, which is parylene only (orange curve) or a bilayer composed of PMMA and parylene (black curve).

**Bode plot for the capacitors**

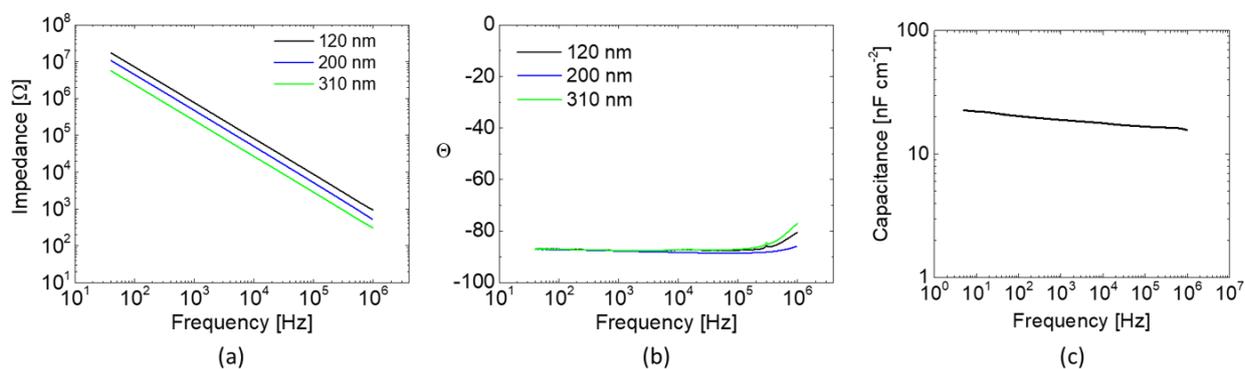

**Figure S2.** Bode plot of the capacitors. Shown are the impedance (a) and the phase (b) as a function of frequency. (c) Capacitance as a function of frequency for capacitor employing a 120 nm thick parylene layer. The capacitance here is measured down to 5 Hz, and the value recorded at the minimum frequency is the one used in order to extract the field effect mobilities.

**Transistors yield – raw data of all the measured transfer curves**

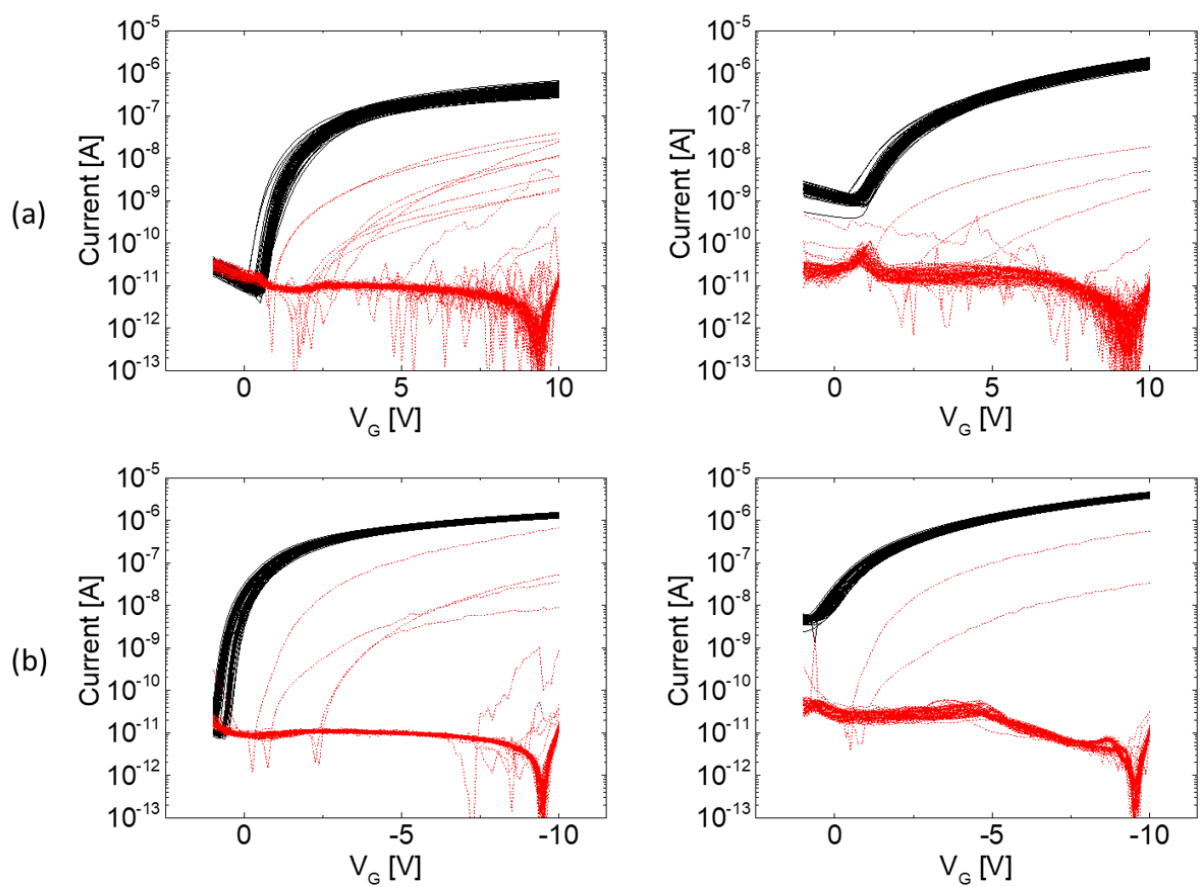

**Figure S3.** Raw data of the transfer curves for the 10x10 transistors array. Data for the n-type (a) and p-type (b) transistors, with linear regime shown on the left (measured at 2 V), and saturation regime on the right (measured at 10 V). Red curves refer to the leakage current.

**Mobility curves**

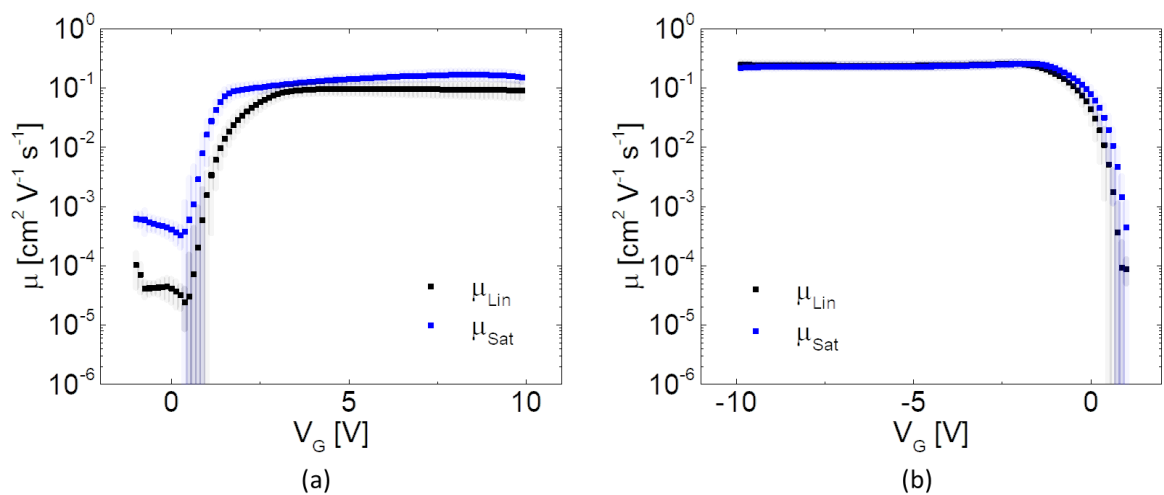

**Figure S4.** Average mobility curves as a function of the gate voltage, together with their standard deviation, for *n*-type (a) and *p*-type (b) devices, in linear (±2 V, black curve) and saturation (±10 V, blue curve) regime.

**Optical micrograph of defective transistors**

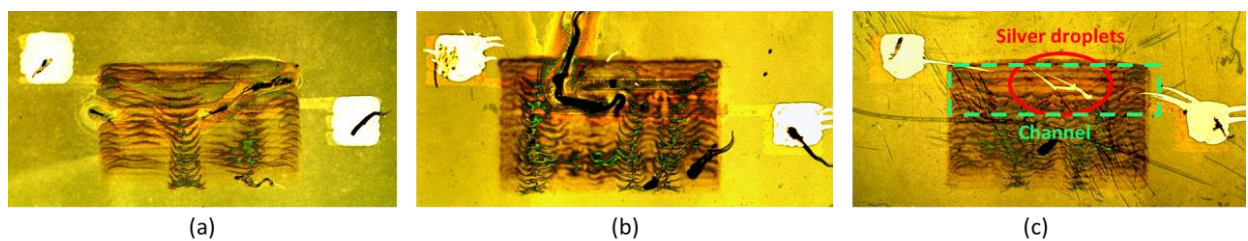

**Figure S5.** Optical micrograph of the defective transistors. (a) and (b) show the two transistors that are not working. In both cases the cause of failure is to be related with the presence of dust particles inside the channel. (c) Example of a transistor with non-ideal leakage, caused by the presence of silver droplets inside the channel.

**Bending test results**

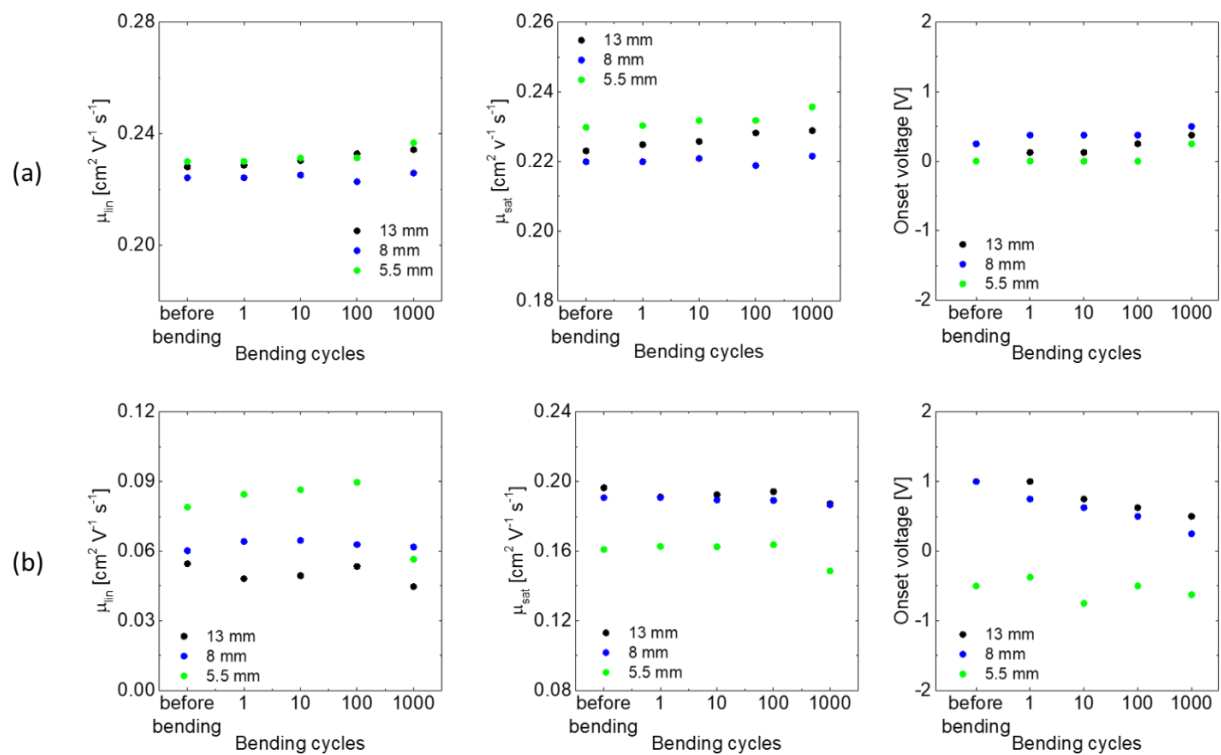

**Figure S6.** Effect of the bending tests on different transistors' figures of merit, for p-type (a) and n-type (b) devices. From left to right, mobility evaluated in linear regime (2 V), mobility in saturation regime (10 V) and onset voltage.

**Performances of the inverter after 4 months storage in nitrogen atmosphere**

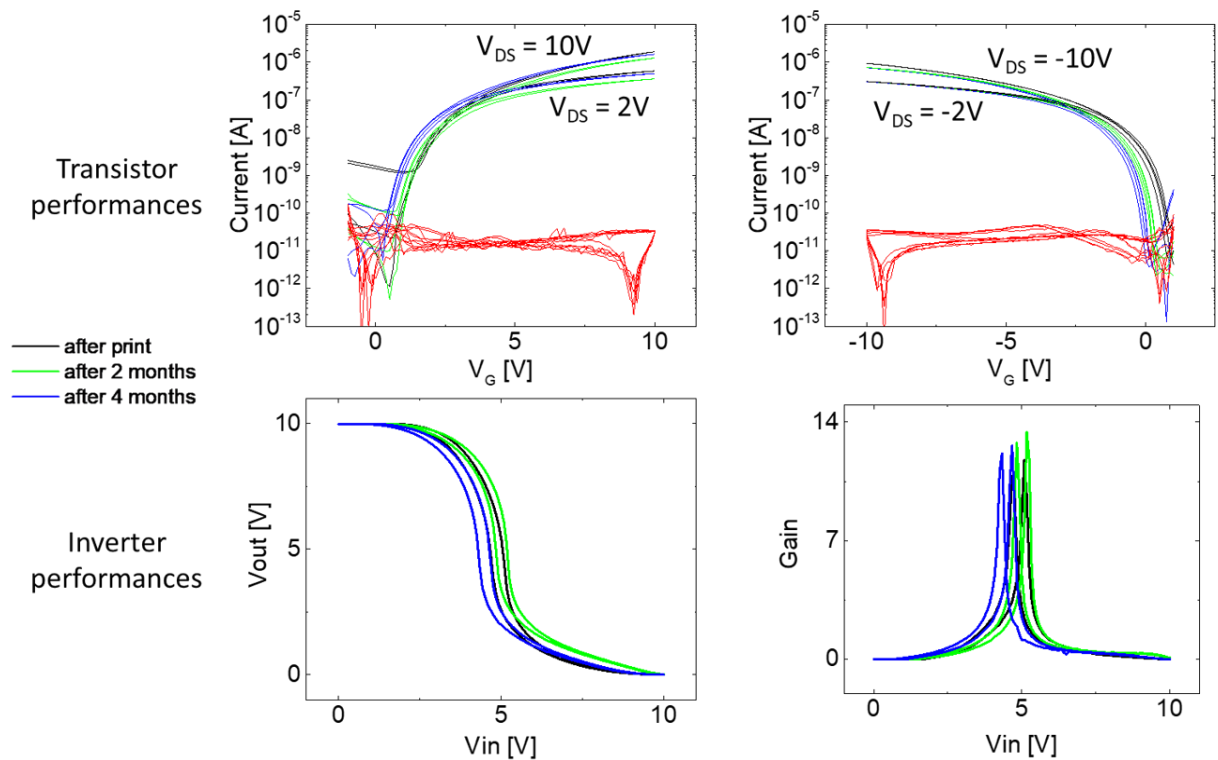

**Figure S7.** Performances of the same n-type transistor (a), p-type transistor (b) and of the same inverter (c-d) right after the device fabrication, after 2 months and after 4 months of storage in nitrogen atmosphere.

**D-Latch**

The D-Latch is a device used to store 1 bit of information. This circuit is composed of two inverters and two transmission gates, working with opposite phases, and it requires only 6 transistors to be implemented with the technology we are using. The operation of the D-latch can be divided into two main phases, the *transparency* and the *hold* phase. During the first, the clock is high and the input signal passes through the circuit and reaches the output, so the latch is said to be in transparent mode. On the contrary, when the clock is low, the device holds stably as output the signal received as input on the falling edge of the clock. Analyzing in details what happens inside the circuit, during the transparency phase, when the clock (CK) is high, and so

CK' is low, the first transmission gate TG1 is acting as short circuit, while TG2 is acting as an open circuit. The input DATA is thus inverted twice, by both the first inverter INV1 and the INV2, and then transmitted to the output node, so the output of the transparency phase is the delayed copy of the input signal. On the contrary, during the hold phase, the clock CK is low and CK' is high, so TG1 is acting as open circuit and TG2 as short, which means that they form a closed loop and the output node corresponds to the input of the inverter INV1. The output is not sensitive to any change in DATA during the opaque phase.

The operation of the two latches DL1 and DL2 of the working D-Flip-Flop presented in this paper is shown in Figure S7.

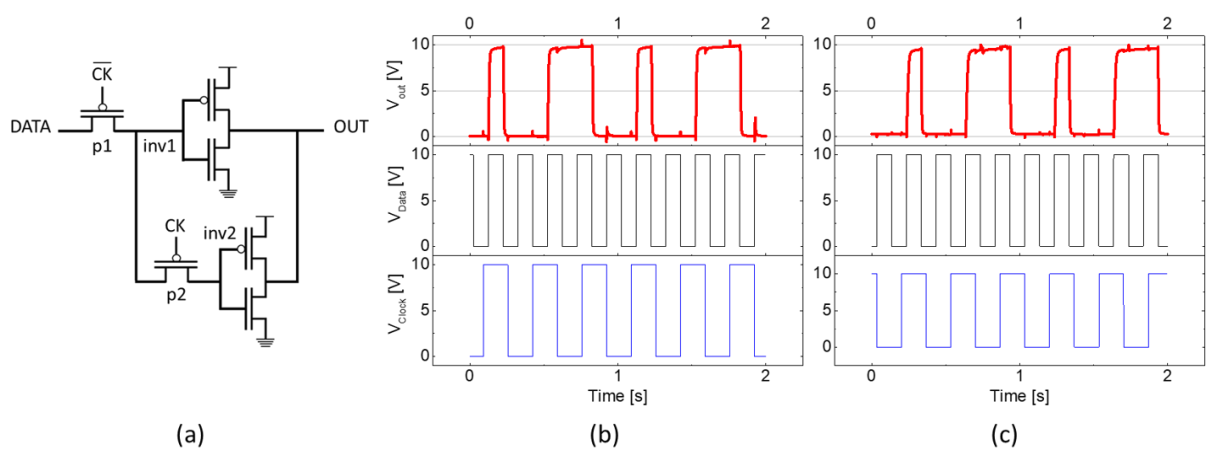

**Figure S8.** (a) Schematic representation of the printed, all-organic D-Latch. Operation of the first (b) and the second (c) D-Latch composing the D-Flip-Flop presented in this paper.

**D-Flip-Flop**

The Master-Slave D-Flip-Flop is one of the basic circuits used to store information. The Master-Slave DFF is implemented by making use of two D-Latch circuits, connected in series and with opposite clock signals, so that when one is in its transparency phase, the other is in opaque phase, and vice versa. The output signal of this device is kept constant all the time and, being a falling-edge triggered device, might change its output value only when the clock signal is switching from high to low. In fact, in that instant, the first D-Latch DL1 goes from its transparency phase to the hold one, sampling the input data and then storing it into its feedback loop, while the second D-Latch DL2 switches from opaque to transparent, making thus possible for the input signal sampled during the falling edge of the clock to reach the output node.

**UV-vis spectra measurement details**

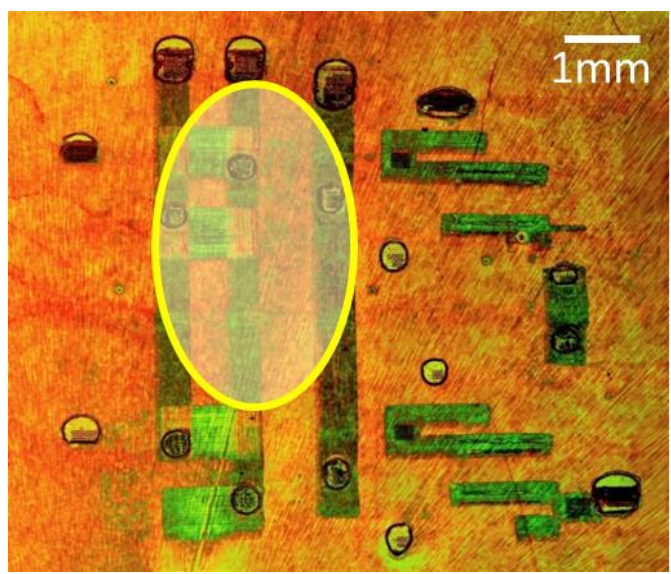

**Figure S9.** Schematic of the spot where the UV-vis spectra has been measured.